\documentclass[conference]{IEEEtran}
\IEEEoverridecommandlockouts

\usepackage{cite}
\usepackage{amsmath,amssymb,amsfonts}
\usepackage{algorithmic}
\usepackage{graphicx}
\usepackage{textcomp}
\usepackage{xcolor}
\usepackage{multirow}
\usepackage{etoolbox}

\def\BibTeX{{\rm B\kern-.05em{\sc i\kern-.025em b}\kern-.08em
    T\kern-.1667em\lower.7ex\hbox{E}\kern-.125emX}}
\begin{document}

\providetoggle{anon}
\settoggle{anon}{false}

\title{\textbf{\textit{Bayesian A/B Testing for Business Decisions}}}
\nottoggle{anon}{
  \author{
    \IEEEauthorblockN{Shafi Kamalbasha and Manuel J. A. Eugster}
    \IEEEauthorblockA{Avira Operations GmBH \& Co. KG \\
    Tettnang, Germany}}
}{
  \author{
    \IEEEauthorblockN{Anonymous}
    \IEEEauthorblockA{Institution \\
    Place}}
}

\maketitle

\begin{abstract}
Controlled experiments (A/B~tests or randomized field experiments) are the
de facto standard to make data-driven decisions when implementing changes and
observing customer responses. The methodology to analyze
such experiments should be easily understandable to stakeholders like product and
marketing managers. Bayesian inference recently gained a lot of popularity and,
in terms of A/B~testing, one key argument is the easy interpretability.
For stakeholders, ``probability to be best'' (with
corresponding credible intervals) provides a natural metric to make
business decisions.
In this paper, we motivate the quintessential questions a business owner
typically has and how to answer them with a Bayesian approach. We present
three experiment scenarios that are common in our company, how they are
modeled in a Bayesian fashion, and how to use the models to draw business
decisions. For each of the scenarios, we present a real-world experiment, the
results and the final business decisions drawn.
\end{abstract}


\section{Introduction}

Controlled experiments (A/B~tests or randomized field experiments) are the de
facto standard to make data-driven decisions; to drive innovation by the
possibility of evaluating new ideas cheaply and quickly. Online experiments
are widely used to evaluate the impact of changes made in marketing
campaigns, software products, or websites.
These days, basically every large technology company continuously
conducts experiments:
Facebook for user interface changes in their News Feed~\cite{bakshy+-2014};
Google states that \textit{``experimentation is practically a mantra; we evaluate
almost every change that potentially affects what our users
experience.''}~\cite{tang+-2010}; and Microsoft for optimizing
Bing ads~\cite{Kohavi+-2013}.
In all cases, the key reason to conduct controlled experiments is the possibility
to establish a causal relationship (with high probability) between a change on
a website or a new feature in a product and changes in visitor or user
behavior~\cite{gupta+-2019}.

\textit{``It's all about data these days. Leaders don't want to make decisions
unless they have evidence.''} states the Harward Business Review in an
2017~article \textit{``A Refresher on A/B Testing''}~\cite{forbes-2017} with the
subtitle \textit{``Spoiler: Many people are doing it wrong.''}
The article stresses the importance of experiments for businesses, but also
the difficulties in communicating the Frequentist~inference-based results
like \textit{significance}, \textit{confidence interval}, and
\textit{$p$-value} to stakeholders like
marketing and product managers. In recent years, Bayesian inference
gained popularity as it has several
advantages over Frequentist inference~\cite{krypotos+-2017}. One key argument
in our business setting is that the result expressed in
\textit{probability to be the best variant} and \textit{expected uplift in
conversion rates} are easier to communicate to stakeholders and, consequently,
it is easier for them to make business decisions.

In our company, we are continuously running A/B tests in marketing, website, and
product development. In this paper we show the different experimentation
scenarios and how we model them in terms of Bayesian inference; the paper is
organized as follows.
Section~\ref{sec:exp} discusses the general structure of experiments and the
main business questions to be answered with them.
Section~\ref{sec:bayes} reviews the basics of Bayesian inference and
introduces relevant concepts and distributions needed in this paper.
Section~\ref{sec:models} introduces the three
business scenarios we tackle and the corresponding models.
Section~\ref{sec:decision} describes the methodology to draw business decisions
based on the introduced models.
Section~\ref{sec:examples} presents three real-world cases where we used
this methodology to come to a business decision. We provide empirical
verification of the correctness by comparing the results
to a state-of-the art commercial experimentation tool.
Section~\ref{sec:summary} concludes this paper with a summary and the discussion
on future work. The open source implementation of the methodology is freely
available to the community from \texttt{https://github.com/Avira/prayas}.

\section{Experiments}
\label{sec:exp}

\begin{figure}[htbp]
\centerline{\includegraphics[width=1\columnwidth]{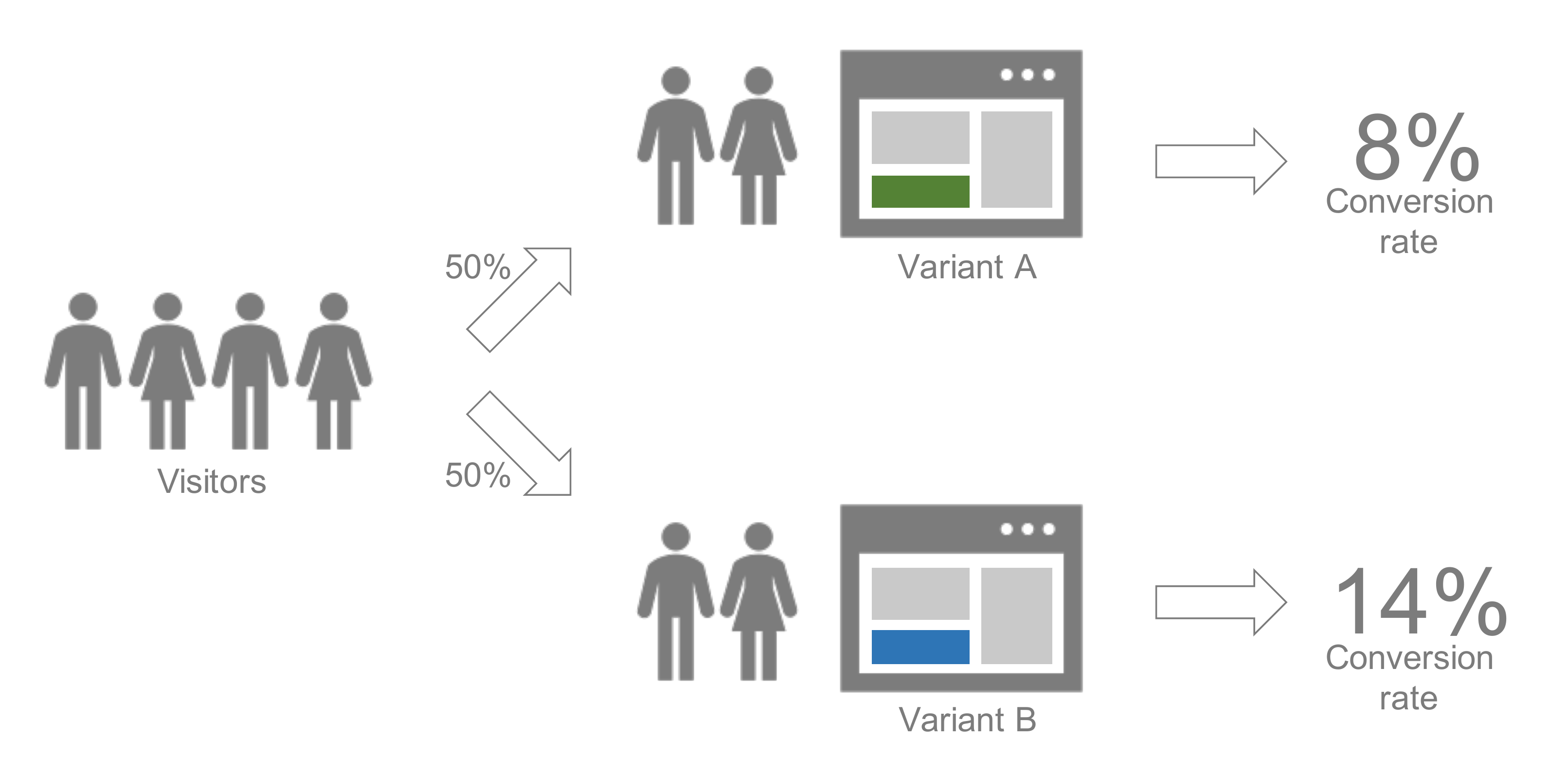}}
\caption{Illustration of a basic experiment with two variants that only differ
in the color of the call-to-action button. The visitors are randomly split into
two groups one assigned to Variant~A, the other one to Variant~B. The goal of
the experiment is to find out the effect of this change on the measure of
interest \textit{conversion rate}.
}
\label{fig:abtest}
\end{figure}

The main objective of a business is to maximize revenue. As described in the
previous section, conducting controlled experiments is an effective way to do so.
Let us take an example of an online e-commerce business with an objective of
increasing its transactions, Fig.~\ref{fig:abtest} illustrates the scenario.
It currently has a green ``purchase'' button on its
webpage and wants to experiment by changing the colour to blue and see if this
change has an effect on the number of transactions. We would call
the page with the green button variant~\textit{A} (the control) and
the one with blue variant~\textit{B} (the treatment).
In this paper, we focus on conversion-based experiments. Therefore, every
experiment has trials and successes. Each visitor to the webpage counts
to a trial and each purchase counts to a success (we also refer to a success
as a conversion). The measure of interests are the uplift in
conversion rate, and revenue-based metrics like average order value and
customer lifetime value.

In the execution of the experiment, each visitor is randomly assigned to
variant~\textit{A} or variant~\textit{B}. Given a valid design and a correct
execution, the only difference between the two variants is the color of the
button; all other factors such as seasonality or moves by the competition occur in
both control and treatment. This means, that the difference in the
measures of interests are either due to the change of the button color or by random
chance~\cite{gupta+-2019}. To determine the ``relevance'' of the difference,
statistical methods are used---either a statistical test in terms of
the Frequentist inference or,
as we motivate here, a model driven approach in terms of Bayesian inference.

In detail, we present a Bayesian approach to answer the quintessential questions
that typically arise in the mind of a business owner:
\begin{enumerate}
\item What is the probability that variant~\textit{B} is better than variant~\textit{A}?
\item How much uplift in conversions will variant~\textit{B} bring?
\item What is the risk of switching to variant~\textit{B} from variant~\textit{A}?
\end{enumerate}
Note that in the course of this paper we define the
approach with experiments executed on websites, with visitors as the
population, the purchase of a product as the success event, and additional
revenue-based metrics as measure of interest. However the
approach is generally applicable, e.g., also in the domain of product testing.

\section{Bayesian statistics}
\label{sec:bayes}

In this section we provide an introduction into relevant
concepts of Bayesian statistics. We mainly follow
\textit{``Bayesian Data Analysis''} by Gelman et. al~\cite{gelman+-2013} and
for specific details we refer to this seminal book.

Bayesian statistics is based on the principle of probability statements and
how to update probabilities after obtaining new data. The \textit{Bayes' theorem}
is the fundamental model describing such updating:
\begin{equation}
P(\lambda | X) = \frac{P(\lambda)P(X | \lambda)}{P(X)}
\end{equation}
It describes the dependency of the \textit{posterior} distribution $P(\lambda | X)$
of a parameter $\lambda$ after seeing data $X$ with the
\textit{prior} distribution $P(\lambda)$ of the parameter $\lambda$,
the \textit{likelihood} $P(X | \lambda)$ of the data $X$ given the
parameter $\lambda$, and the \textit{marginal likelihood} $P(X)$ of the data $X$.
In many cases, also in this paper, we can ignore the marginal likelihood and
work with $P(\lambda | X) \propto P(\lambda)P(X | \lambda)$.
With this, the primary task in developing Bayesian models for specific
applications is to define proper models for the prior and the likelihood.

A very important concept in Bayesian statistics is \textit{conjugacy}, meaning
that the posterior distribution $P(\lambda | X)$ is in the same probability
distribution family as the prior distribution $P(\lambda)$. The prior is then
called a \textit{conjugate prior}. This allows us to
write the posterior in a closed-form expression which is mathematically as
well as computationally convenient.

\subsection{Relevant distributions}

\begin{figure}[htbp]
\centerline{\includegraphics[width=1\columnwidth]{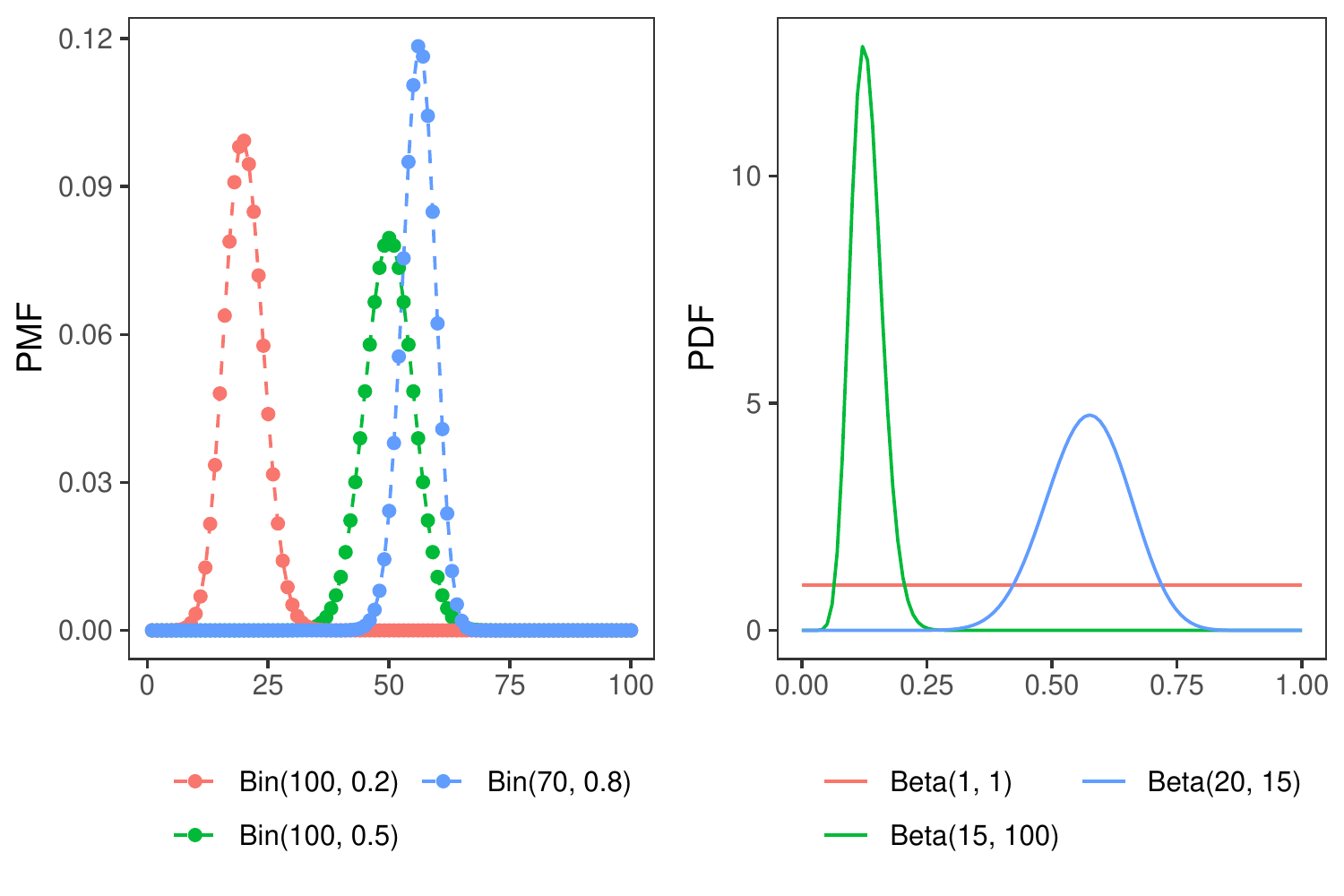}}
\caption{(left)~Probability mass function of the Binomial distribution
$Bin(n, p)$ for different values of $n$ and $p$. This distribution is a
discrete distribution, the line between the points is just to increase
readability.
(right)~Probability density function of the Beta distribution
$Beta(a, b)$ for different values of $a$ and $b$.
}
\label{fig:distr}
\end{figure}

To describe the models in this paper we need six distributions that we
briefly introduce.

The \textit{binomial} distribution $Bin(n, p)$ is a discrete distribution defined on
the interval $\{0, \ldots, n\}$. It describes the number of successes in
$n \in \mathbb{N}$ trials with success probability $p \in [0, 1]$ in each trial.
Fig.~\ref{fig:distr}~(left) shows the probability mass function
for different parameter values.

The \textit{Beta} distribution $Beta(a, b)$ is a continous
distribution defined on the interval $[0, 1]$. The parameters $a$ and $b$ define
the shape of the distribution, e.g., $a = b = 1$ gives the standard uniform
distribution. Fig.~\ref{fig:distr}~(right) shows the probability density
function for different parameter values. The Beta distribution is the conjugate
prior for the binomial distribution.

The \textit{multinomial} distribution $Mult(n, \underbar{p})$ is the generalization of the
binomial distribution. Compared to the binomial distribution, it describes the
number of successes in $n \in \mathbb{N}$ trials for $k > 0$ options. The
distribution gives the probability of any particular combination of numbers of successes for the
$k$ options with the fixed
success probability $\underbar{p} = [p_1, \ldots, p_k]$ and $\sum_{i}^k p_i = 1$.

The \textit{Dirichlet} distribution $Dir(\underbar{a})$ is the generalization of the Beta
distribution for $k \geq 2$ options. The parameter $\underbar{a}$ is defined as
$\underbar{a} = [a_1, \ldots, a_k]$ with $a_i > 0$. The Dirichlet distribution
is the conjugate prior for the multinomial distribution.

The \textit{exponential} distribution $Exp(\theta)$ is a continuous distribution
defined on the interval $[0, \inf]$ with the parameter $\theta > 0$ called
the scale (another common parametrization is with the rate parameter $\theta^{-1}$).
We use the exponential distribution to model revenue, i.e., the
higher the price of a product the lower the probability to be purchased.

The \textit{Gamma} distribution $Gamma(\alpha, \beta)$ is a generalization of the
exponential distribution. The parameters $\alpha > 0$ define the shape and
$\beta > 0$ the scale of the distribution; with the parameter $\alpha = 1$ it
is an exponential distribution with $\theta = \beta$.
The Gamma distribution is the conjugate prior for the exponential distribution.

\section{Experiment scenarios and models}
\label{sec:models}

In our company, majority of experiments fall into one of the three scenarios:
1)~compare variants where the visitor has one option to choose,
2)~compare variants where the visitor has multiple options to choose from, and
3)~compare variants where the visitor has multiple options to choose from but
we only observe the overall success. In the following we describe a model
for each scenario.

\subsection{One option model}
\label{sec:one}

\begin{figure}[htbp]
\centerline{\fbox{
\nottoggle{anon}{
  \includegraphics[width=0.95\columnwidth]{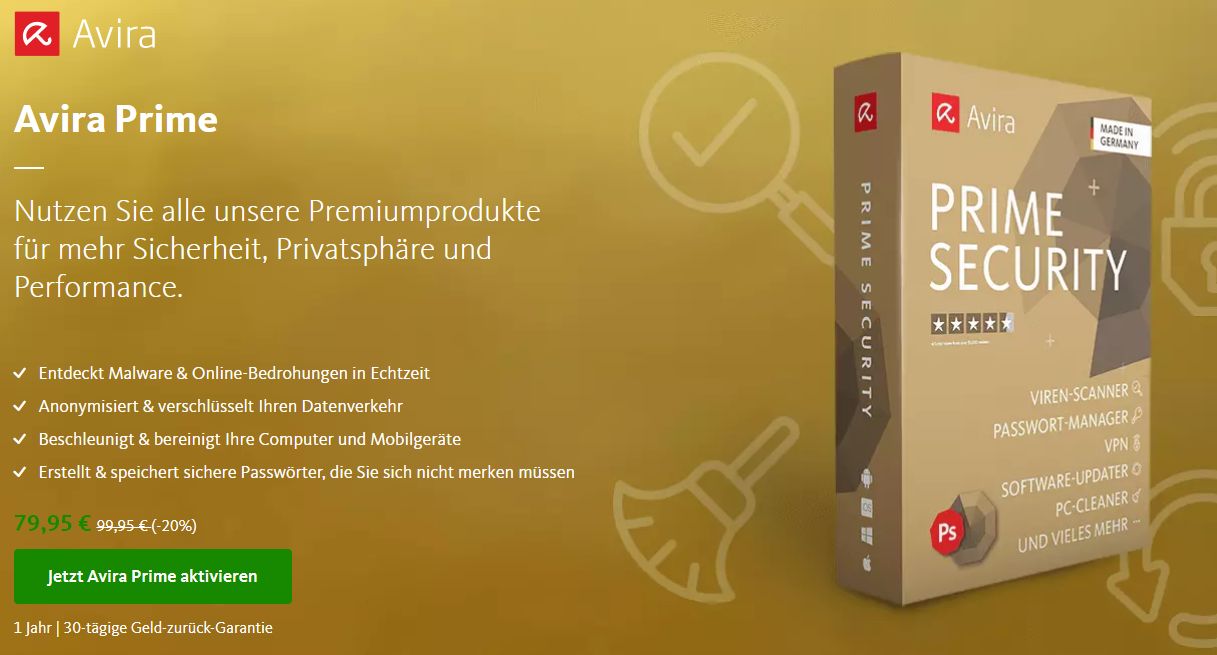}
}{
  \includegraphics[width=0.95\columnwidth]{fig-oneoption-varianta-anon.png}
}
}}
\caption{Example of one variant in case of a one option model experiment. The website
visitor has only the option to purchase one product. The purchase
of the presented product is the event of interest for this variant.}
\label{fig:oneoption-varianta}
\end{figure}

The first model we describe is applicable in the basic scenario of experiments
described in Section~\ref{sec:exp} and illustrated in Fig.~\ref{fig:abtest};
Fig.~\ref{fig:oneoption-varianta} shows a real-world variant of one of our
experiments where the visitor has only the option of purchasing one product.
One option scenarios are formalized as follows.

The experiment consists of variants $i$ with $i = 1, \ldots, V$ where $V$ is the
total number of variants. Each variant displays only one option to the visitor.
For a given variant $i$, $N_i$ is the total number
of visitors that saw variant $i$ (trials), and $C_i$ the number of conversions
(successes). Success is defined by the occurrence of the event
of interest, for example, the visitor clicks a button click or
purchases a product. A success also has a value $v_i$ such as revenue or
customer lifetime value associated and can be different for every variant.
In a frequentist approach, the conversion rate $\lambda_i$ is the point
estimate $\frac{C_i}{N_i}$, the total revenue earned is
$N_i \cdot \lambda_i \cdot v_i$, and the revenue earned per visitor per variant
is $\lambda_i \cdot v_i$.

Executing the experiment results in collected data for each variant,
$X_i = \{x_{i1}, \ldots, x_{iN_i}\}$ with $x_{ij} \in \{0, 1\}$,
$j = 1, \ldots, N_i$ and $0$ indicates a non-conversion and $1$ a conversion.
Compared to the frequentist approach explained above, we model the data
generating process and the distribution of the conversion
rate~\cite{davidsonpilon-2015}.
Since the conversions are Boolean-valued, the sequence of
conversions $X_i$ follows the binomial distribution and the generative process
of the data is:
\begin{equation}
P(X_i | \lambda_i) = Bin(N_i, \lambda_i)
\end{equation}
The prior distribution about the conversion rate is modeled as
\begin{equation}
P(\lambda_i) = Beta(a_i, b_i) \text{,}
\end{equation}
with the hyperparameters $a_i$ and $b_i$ defined in a way to represent our
prior belief. Fig.~\ref{fig:distr}~(right) shows different prior beliefs about
the conversion rate: with $a_i = b_i = 1$ we have an uniformed prior and all
conversion rates are equally likely; with $a_i = 15$ and $b_i = 100$ we define
a prior where the conversion rate is around $0.125$ with small variance, meaning
that we have a strong prior knowledge; and with $a_i = 20$ and $b_i = 15$ the
conversion rate is around $0.6$ with wider variance, indicating that we
have only weaker knowledge about the conversion rate.

To compute the posterior, we leverage the fact that the conjugate prior for a
distribution belonging to binomial family is in the beta family, and
therefore the posterior distribution is the same as the prior distribution
with updated parameter values~\cite{gelman+-2013}:
\begin{equation}
\label{eq:post-one}
P(\lambda_i | X_i) = Beta(a_i + C_i, b_i + (N_i - C_i))
\end{equation}
The posterior $P(\lambda_i | X_i)$ gives the probability distribution of all possible
values of $\lambda_i$ given the evidence $X_i$. We can draw $n$ random samples
from the posterior to obtain a set $Y_i = [y_{i1}, \ldots, y_{in}]$ of possible
values the conversion rate $\lambda_i$ can take. The possible values for the
revenue (or any other value $v_i > 0$) per visitor is defined as
$Z_i = Y_i \cdot v_i$.
There are cases where businesses incur loss for each
non-conversion. One example is the pay-per-click model
followed in online advertising industry where businesses must pay the
ad platform for each click generated regardless of conversion. The
above model can be adjusted to penalize the loss $l_i$ of each
non-conversion as $Z_i = Y_i \cdot (v_i - l_i) + (1 - Y_i) \cdot (-l_i)$.

\begin{figure}[htbp]
\centerline{\includegraphics[width=1\columnwidth]{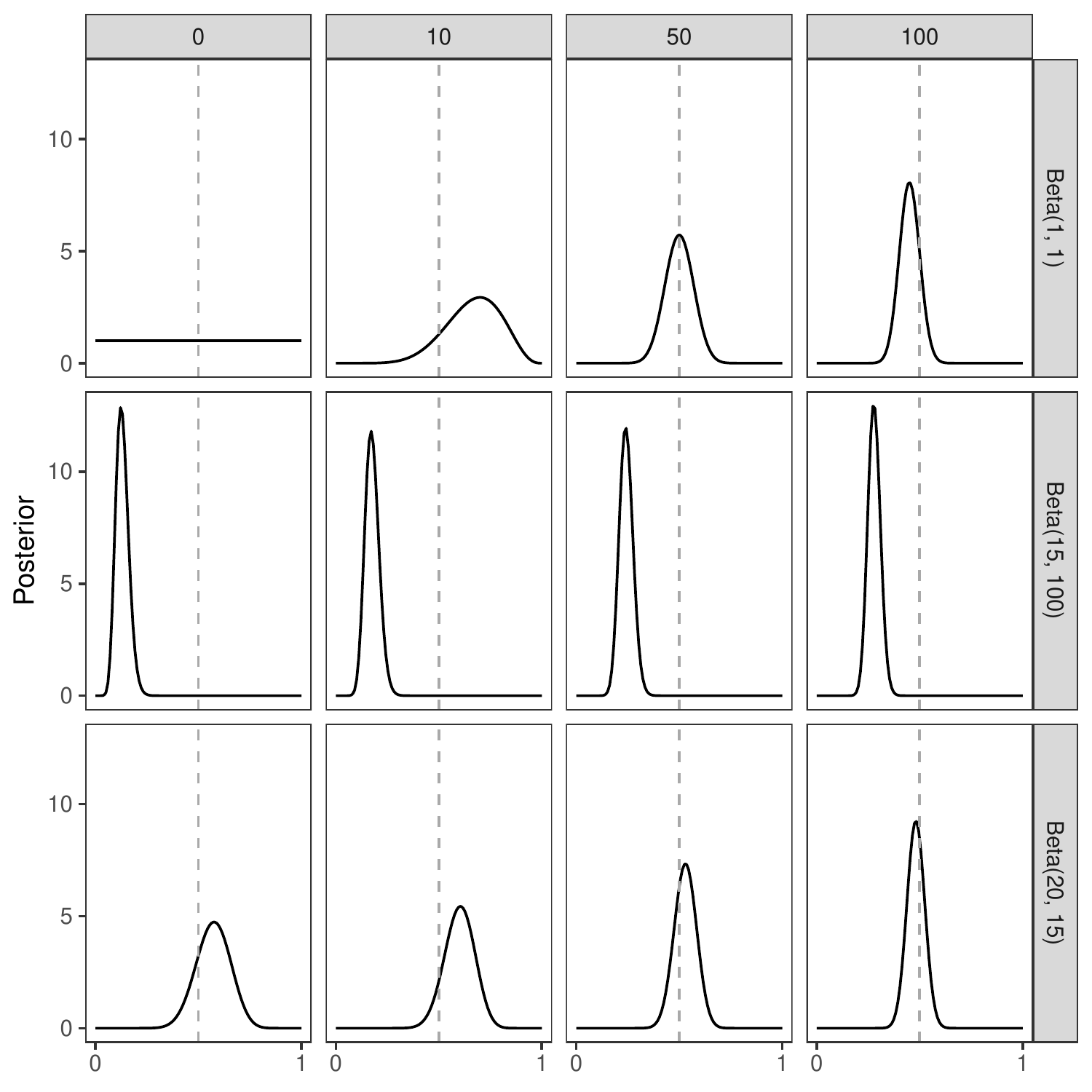}}
\caption{Illustration of the Bayesian updating. For simulated data, the
posterior is shown for the three priors defined in
Fig.~\ref{fig:distr}~(right) after seeing 10, 50, and 100 samples; column~0
shows the corresponding prior. The dashed vertical line indicates the true
conversion rate $0.5$.}
\label{fig:oneoption-sim}
\end{figure}

\paragraph{Simulation of the Bayesian updating} To illustrate the Bayesian
updating from prior to posterior when gathering new data, we simulate a
one option experiment and take a closer lock at Variant $1$. For the
data $X_1$ we draw $N_1 = 100$ samples from a binomial distribution with the
conversion rate $\lambda_1 = 0.5$. Starting from the priors shown in
Fig.~\ref{fig:distr}~(right), we look at the posteriors $Y_1$ after seeing
the first 10, 50, and all 100 samples. Fig.~\ref{fig:oneoption-sim} shows the
simulation. In the first row, we start with an uninformative prior, and after
seeing all data the posterior is near to the true conversion rate. In the
second row, we start with a strong prior knowledge of the conversion rate being
around $0.125$; here 100 samples are not enough to estimate the true conversion
rate. In the third row, we start with weaker prior knowledge of the conversion
rate being around $0.6$, and in this case, the posterior is near the true
conversion rate after seeing all samples.

\subsection{Multi-options model}
\label{sec:multi}

\begin{figure}[htbp]
\centerline{\fbox{
\nottoggle{anon}{
  \includegraphics[width=0.95\columnwidth]{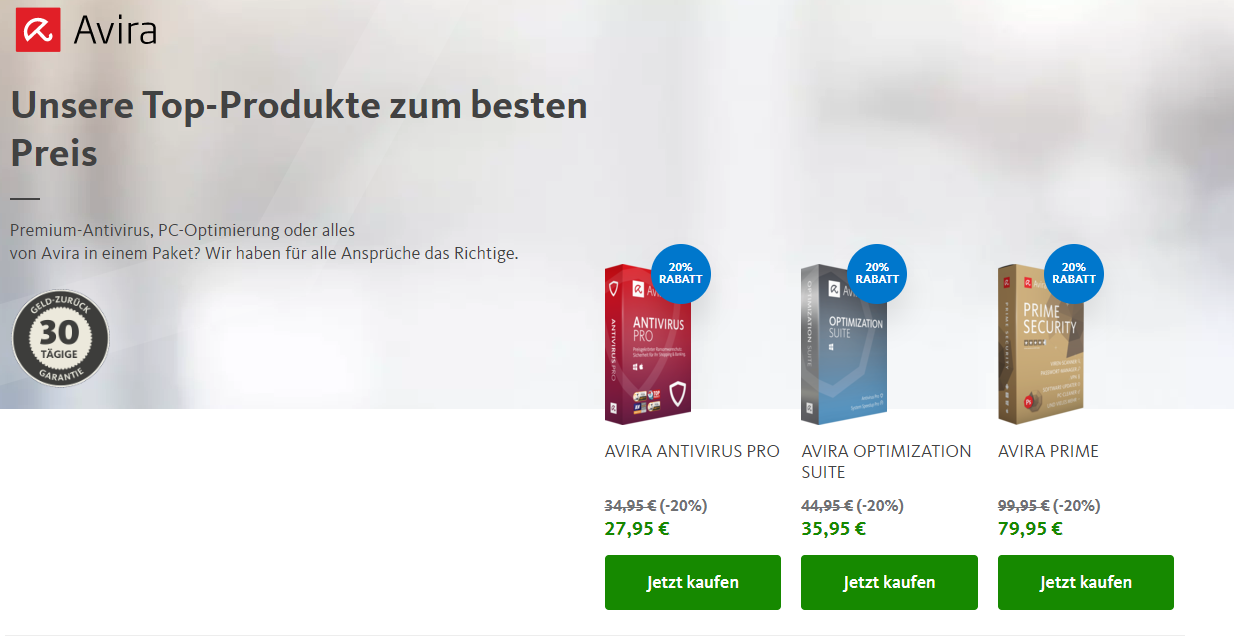}
}{
  \includegraphics[width=0.95\columnwidth]{fig-multioptions-varianta-anon.png}
}
}}
\caption{Example of one variant in case of the multi-options model. The website
visitor has multiple options to select, in this case to purchase three different
products. The purchases of the presented products are the events of interest
for this variant.}
\label{fig:multiptions-varianta}
\end{figure}

The second model we describe is the generalization of Scenario~1.
Scenario~1 assumes that there can be only one possible event of interest
in each variant. However, in real world, it oft is too strict a restriction.
For example, many online websites showcase multiple product purchase buttons in
a single page. Fig.~\ref{fig:multiptions-varianta} shows a variant of one of
our experiments where visitors can choose from three different product options.
Multi-options scenarios are formalized as follows.

Each variant $i$ displays $K_i$ different options, from which the visitor can
chose either one of the options or none. The number of conversions is given by
$C_i = [c_{i}^{1}, \ldots, c_{i}^{K_i}]$ successes for each option per variant.
The revenue for each success value is $v_i = [v_{i}^{1}, \ldots, v_{i}^{K_i}]$
since each of the multiple options can have a different value (e.g., due to
different prices per product). The conversion rate is defined as
$\lambda_i = [\lambda_{i}^{1}, \ldots, \lambda_{i}^{K_i+1}]$
with $\lambda_{i}^{1}, \ldots, \lambda_{i}^{K_i}$ the probabilities of
choosing individual $K_i$ options and $\lambda_{i}^{K_i+1}$ the
probability of choosing none of the options. Since each visitor can only choose
from one of these $K_i+1$ options, they are mutually exclusive and
therefore for every variant $i$ it holds that
$\sum_{l=1}^{K_i+1} {\lambda_i^{l}} = 1$.


Executing the experiment results in collected data for each variant,
$X_i = \{x_{i1}, \ldots, x_{iN_i}\}$
with $x_{ij} =  [x_{ij}^{1}, \ldots, x_{ij}^{K_i}]$
and $x_{ij}^{l} \in \{0, 1\}$ indicating which option the visitor chose.
Since we consider multiple events within a variant, the sequence of conversions
follows a multinomial distribution and the generative process of the
data is~\cite{davidsonpilon-2015}:
\begin{equation}
P(X_i | \lambda_i) = Mult(N_i, \lambda_i)
\end{equation}
The prior distribution of the conversion rate is modeled as
\begin{equation}
P(\lambda_i) = Dir(a^1, \ldots, a^{K_i + 1}) \text{,}
\end{equation}
where $a_\cdot$ are the hyperparameters.

Once again, leveraging the conjugacy relationship between multinomial and
Dirichlet distributions, we compute the posterior distribution with updated
parameter values as~\cite{gelman+-2013}:
\begin{equation}
\label{eq:post-multi}
\begin{split}
P(\lambda_i &| X_i) = \\
& Dir(a^1 + c^1, \ldots, a^{K+1} + (N_i - (c^1 + \ldots + c^K))
\end{split}
\end{equation}
From this posterior distribution of conversion rates, we can
draw $n$ random samples $Y_i = [y_{i1}, \ldots, y_{in}]$
where each element $y_{ij}$ is the overall conversion rate over all options
available in the Variant~$i$. 
Similarly to the one option model, we can penalize the non-conversions, and
we can compute a revenue-based metric as the sum of
element-wise multiplication of the values $v_i > 0$ and $y_{ij}$.

\subsection{Aggregated model}
\label{sec:aggr}

This scenario is a special case of the scenarios 1 and 2 where only the
aggregated revenue and the sum of conversion per variant are observed.
This can occur because
of the data generation process or simply because
there are too many options within a variant making it cumbersome to
model it under Scenario~2. Some of our experiments have up to 81 different
options the visitor can chose from, due to the different products listed together
with the options for license runtimes and the number of devices on which
the product can be installed.
This aggregated scenario is formalized as follows.

For a given variant $i$, there are $K_i$ unknown options. $N_i$ is the number
of visitors and $C_i$ is the number of conversions defined as in the
multi-option model.
The individual revenue per option $v_i = [v_{i}^{1}, \ldots, v_{i}^{K_i}]$
is unknown. Executing the experiment results in collected data $C_i$, $N_i$ and
the aggregated revenue $s_i$ over all $C_i$ successes and
implicitly over all unknown $K_i$ options, i.e., $s_i = \sum_{j=1}^{C_i} s_{ij}$
with $s_{ij}$ the revenue of each success $j$.

The posterior for the conversion rate $\lambda_i$ can be obtained by following
the one option model and the therein defined posterior in
Equation~\ref{eq:post-one}. To get an estimate of the revenue per visitor
$v_i$ we model the average revenue per visitor $\bar{v}_i$ given the
observed aggregated revenue $s_i$. For that we
assume that $s_{ij}$ follows an exponential distribution
\begin{equation}
P(s_{ij} | \bar{v}_i) = Exp(\bar{v}_i)\text{,}
\end{equation}
where $\bar{v}_i$ is the scale parameter of the distribution~\cite{vwo-2015}.
The exponential distribution means that lower revenue has
more probability of occurrence than higher revenue. This assumption fits our
observation of the visitors' money spent curve on our website.
The prior distribution of $\bar{v}_i$ is modeled as
\begin{equation}
P(\bar{v}_i) = Gamma(\alpha_i, \beta_i)
\end{equation}
Taking advantage of the conjugacy relationship between exponential and
gamma distributions, we compute the posterior distribution as
\begin{equation}
P(\bar{v}_i | s_i) = Gamma(\alpha + C_i, \beta_i + s_i)
\end{equation}
The expected value per visitor per variant is $\lambda_i * \bar{v}_i$.
Since we have the posterior distributions for both, $\lambda_i$ and
$\bar{v}_i$, we draw $n$ random samples from $P(\lambda_i | X_i)$
for the conversion rate $Y_i$, $n$ random samples from
$P(\bar{v}_i | s_i)$ for the revenue.

\section{Decision making}
\label{sec:decision}

Every experiment has one measure of interest defined beforehand, such
as conversion rate or revenue per visitor, on which the variants are
evaluated. In Section~\ref{sec:models}, we described a way of
modelling the measures
of interest and obtaining the samples from the posterior distribution.
Here, we describe the method of comparing different posterior samples
in order to answer the motivating questions from Section~\ref{sec:exp}.

\subsection{Probability to be the best}
After running an experiment with multiple variants, we want to know which
variant is the most favorable and should be implemented. Given a set of
posterior samples $Y_i$ of the measure of interest from $i$ different
variants, the probability to be the best is defined as the probability
that a variant has higher measure in comparison with all other
variants. The probability that $Y_1$ is better than $Y_2$ is the
mean of:
\begin{equation}
P(Y_1 > Y_2) = [y_{1j} > y_{2j} \; | \; i \text{ and } j \in  \{1, \ldots , n\}]
\end{equation}
The extension to more than two variants is to simply compare $Y_1$ against
all others. To find the winner variant we compute all combinations and
select the one with the highest probability.

\subsection{Expected uplift}
The second motivating question we want to answer, once we have the
best variant, is to determine the increase in measure we expect after
its implementation.
Given set of posterior samples $Y_i$, the
expected uplift of choosing $Y_1$ over $Y_2$ is defined as the mean (or the
credible interval) of the percentage increase:
\begin{equation}
U(Y_1, Y_2) = {[\frac{y_{1j} - y_{2j}}{y_{2j}} \; | \; i \text{ and } j \in \{1, \ldots , n\}]}
\end{equation}
As we are interested in the expected uplift compared to the control variant,
we typically only compute this for all treatment variants against the
control variant. However, the equation can be extended to
compare every variant $Y_i$ against each other.

\subsection{Expected loss}
Suppose Variant~1 has the highest probability to be the best but smaller than 1.
Then, there is
still a chance that the other variant is the true best performing one. In such
a case we want to know the risk of implementing Variant~1.
Given set of posterior samples $Y_i$, the
expected loss when choosing $Y_1$ over $Y_2$ is the mean (or the
credible interval) of:
\begin{equation}
L(Y_1, Y_2) = [\max{(\frac{y_{2j} - y_{1j}}{y_{1j}}, 0)} \; |  \; i \text{ and } j \in  \{1, .. , n\}]
\end{equation}
Similar to the expected uplift, expected loss is typically calculated between the
the treatment variant and the control.

\section{Business cases}
\label{sec:examples}

In this section we illustrate the application of Bayesian inference and
decision making while experimenting with different discounts on
our product prices.

\subsection{Single product discount test}

\begin{figure}[htbp]
\centerline{\includegraphics[width=1\columnwidth]{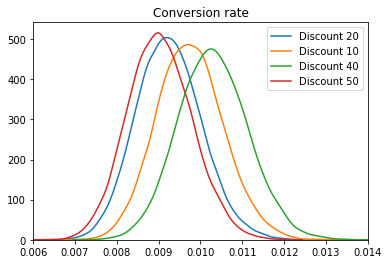}}
\caption{Posteriors for the conversion rate for each variant from the one
option discount experiment. The graph shows the probability density function
with $x$-axis the conversion rate.
}
\label{fig:exp-1}
\end{figure}

\begin{table}[htbp]
\caption{Decision metrics for the one product discount experiment. Baseline (BL)
is the ``Discount 20'' variant. }
\centering
\begin{tabular}{rccccc}
  \hline
  Variant & \multicolumn{2}{c}{Probability to} & \multicolumn{2}{c}{Improvement} & Loss from\\
  & be best & beat BL & Mean & CI95 & BL \\
  \hline
  Discount 20 & 0.10 & &  &   \\
  Discount 10 & 0.26 & 0.68 & 0.06 & [-0.16, 0.33] & 0.02 \\
  Discount 40 & 0.58 & 0.83 & 0.12 & [-0.11, 0.40] & 0.01 \\
  Discount 50 & 0.06 & 0.42 & -0.02 & [-0.23, 0.24] & 0.06 \\
  \hline
\end{tabular}
\label{tab:one-model}
\end{table}

\begin{figure}[htbp]
\centerline{\includegraphics[width=1\columnwidth]{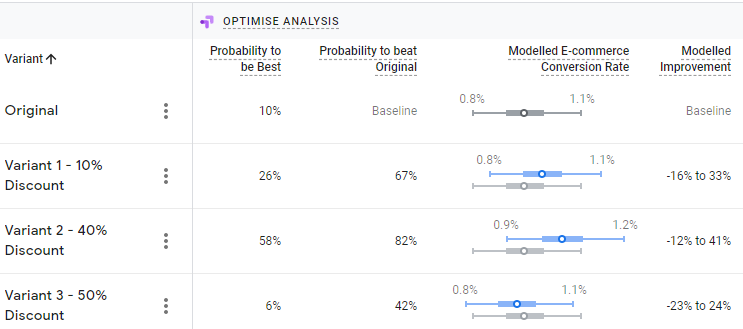}}
\caption{Decision metrics for the one product discount experiment
delivered by Google Optimize and computed with a black-box Bayesian
model~\cite{optimize}.
The results match with our model and the metrics shown in
Table~\ref{tab:one-model}.}
\label{fig:exp-1-optimise}
\end{figure}

We ran an experiment on our main product page, which displayed
20\% discount (see Fig.~\ref{fig:oneoption-varianta}),
with the objective of measuring the effect on
conversion rate by changing the discount. We chose the three treatment
variants to display 10\%, 40\% and 50\% discount. Technically we ran the
experiment with Google Optimize, and we use this experiment to validate our
approach against a industry standard tool. Google Optimize also uses
Bayesian inference for analysis of results, however their concrete models are,
to the best of our knowledge, unknown~\cite{optimize}.

Since there was only one option on the page we used our one option
model, described in Section~\ref{sec:one}, for Bayesian inference
and decision making. After running the
experiment for 53 days, we gathered the following data:
for the variants ``Discount 20\%'', ``Discount 10\%'', ``Discount 40\%'',
and ``Discount 50\%'', the total conversions were 139, 147, 149 and
134 respectively, and the total visitors were 15144, 15176, 14553
and 14948 respectively.

Fig.~\ref{fig:exp-1} shows the posteriors of
conversion rates for all the four variants obtained from the model
using an uninformed prior.
In order to make decisions based on the posteriors, we use the
equations defined in section Section~\ref{sec:decision} to find the probability
to be best,
probability to beat the baseline (Discount 20\%), expected uplift from the baseline,
and the expected loss for each variant. The result in Table~\ref{tab:one-model}
show that he variant ``Discount 40\%'' had the highest probability to
be best (58\%), the
highest average improvement (12\%) and the lowest expected loss (1\%).
Hence, we decided to display 40\% discount instead of 20\% on the
product page.

As a comparison we show the results from Google Optimize in
Fig.~\ref{fig:exp-1-optimise}. All the metrics from Google Optimize and
our model are identical up to some insignificant rounding differences.

\subsection{Multi-product discount test}

\begin{figure*}[htbp]
\begin{center}
\includegraphics[width=.3\textwidth]{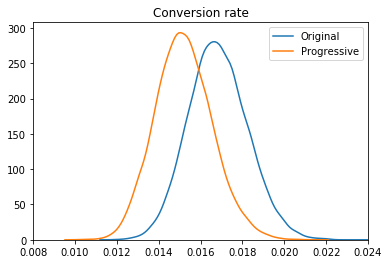}
\includegraphics[width=.3\textwidth]{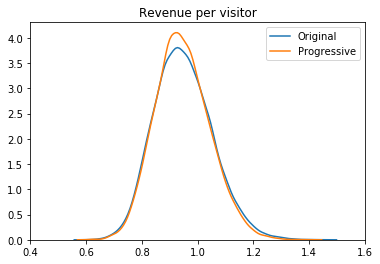}
\includegraphics[width=.3\textwidth]{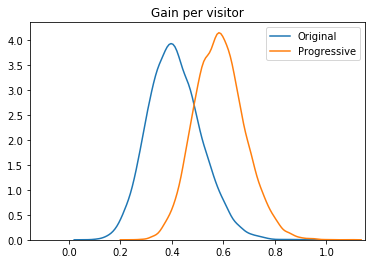}\\
\includegraphics[width=.3\textwidth]{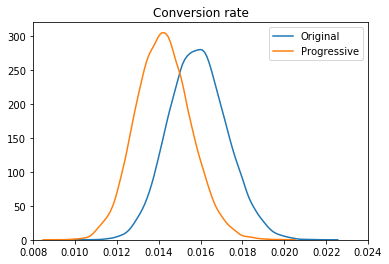}
\includegraphics[width=.3\textwidth]{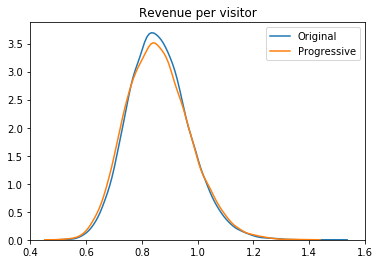}
\includegraphics[width=.3\textwidth]{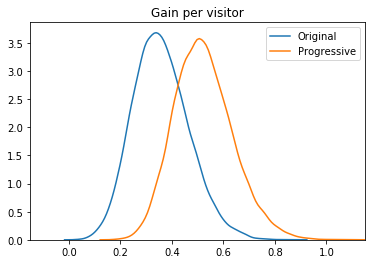}
\end{center}
\caption{Posteriors of the metrics ``conversion rate'', ``revenue'', and
``gain'' for the (top row)~multi-product discount test and
the (bottom row)~multi-product discount test with aggregated data.
The graph shows the probability density functions.
}
\label{fig:exp-2}
\end{figure*}

\begin{table}[htbp]
\caption{Decision metrics for the multi-product discount experiment.
Baseline (BL) is the ``Original'' variant. Probability to be best and
to beat BL are equal as there is only on treatment variant.}
\centering
\begin{tabular}{rccccc}
  \hline
  Variant & \multicolumn{2}{c}{Probability to} & \multicolumn{2}{c}{Improvement} & Loss from\\
  & be best & beat BL & Mean & CI95 & BL \\
  \hline
  \multicolumn{6}{l}{\textit{Conversions:}}\\
  Original & 0.80 & & & &  \\
  Progressive & 0.20 & 0.20 & -0.10 & [-0.32, 0.20] & 0.11 \\
  \hline
  \multicolumn{6}{l}{\textit{Revenue:}}\\
  Original & 0.51 & & & &  \\
  Progressive & 0.49 & 0.49 & -0.01 & [-0.3, 0.41] & 0.06 \\
  \hline
  \multicolumn{6}{l}{\textit{Gain:}}\\
  Original & 0.11 & & & & \\
  Progressive & 0.89 & 0.89 & 0.53 & [-0.19, 1.75] & 0.02 \\
  \hline
\end{tabular}
\label{tab:multi-model}
\end{table}

The second discount experiment was run on our Microsoft Bing Ads landing page,
which displayed three products, each with three different license
options, with 20\% discount each (see Fig.~\ref{fig:multiptions-varianta}).
Since visitors land on
this webpage after clicking our ad, we incur a cost. The
treatment displayed a progressive discount of 0\%, 15\% and 30\%,
in order to nudge the visitors towards our premium product~\cite{thaler+-2008}.
The objective was to measure effect on gain per visitor given as revenue per
visitor minus the cost per visitor.

Since there were total of nine options on each variant, we used our
multi-option scenario described in Section~\ref{sec:multi}.
After running the experiment for 58 days, we
collected the following data: for the variants ``Original'' and
``Progressive'', the total conversions were [50, 5, 5, 28, 7, 5, 20, 1, 6]
and [28, 3, 6, 30, 6, 5, 27, 6, 3], the total visitors were 8067 and
8082, and the revenue for each option per variant was
[27.95, 47.95, 63.95, 35.95, 63.95, 79.95, 79.95, 151.95, 223.95] and
[34.95, 59.95, 79.95, 37.95, 67.95, 84.95, 69.95, 132.95, 195.95]
respectively. The Bing ad cost per click was slightly higher (by 15
cents) for the original variant.

Fig.~\ref{fig:exp-2}~(top row) shows the posteriors of conversion rates,
revenue per visitor and gain per visitor for both the variants, using an
uniformed prior and Table~\ref{tab:multi-model}
shows the result of the experiment. The progressive variant has lower
conversion rates than the original variant (-10\% on average) but the
revenue per visitor is almost the same for both variants since the
probability to be best is almost 50\% for both variants; both variants
are equally likely to be the best. The progressive variant has less
conversions for the lower priced option compared to the original, but
makes up for the lost revenue with higher conversions on the premium
product. For the metric gain per visitor, progressive discount
had the highest probability to be best with a low expected loss of 2\%.
Hence the progressive variant was used in production after the
experiment.

\subsection{Multi-product discount test with aggregated data}

\begin{table}[htbp]
\caption{Decision metrics for the multi-product discount experiment with
aggregated data. Baseline (BL) is the ``Original'' variant.}
\centering
\begin{tabular}{rccccc}
  \hline
  Variant & \multicolumn{2}{c}{Probability to} & \multicolumn{2}{c}{Improvement} & Loss from\\
  & be best & beat BL & Mean & CI95 & BL \\
  \hline
  \multicolumn{6}{l}{\textit{Conversions:}}\\
  Original & 0.80 & & & &  \\
  Progressive & 0.20 & 0.20 & -0.10 & [-0.30, 0.15] & 0.12 \\
  \hline
  \multicolumn{6}{l}{\textit{Revenue:}}\\
  Original & 0.51 & & & &  \\
  Progressive & 0.49 & 0.49 & 0.01 & [-0.3, 0.41] & 0.07 \\
  \hline
  \multicolumn{6}{l}{\textit{Gain:}}\\
  Original & 0.14 & & & & \\
  Progressive & 0.86 & 0.86 & 0.62 & [-0.27, 2.43] & 0.03 \\
  \hline
\end{tabular}
\label{tab:agg-model}
\end{table}

Here we illustrate that the previous example can be modeled under the
aggregated model scenario described in Section~\ref{sec:aggr} to obtain similar
results. We suppose that we only obtained the totals conversions, total
visitors and the overall revenue for the two variants from the above
experiment: 127 and 114 (sum of conversions from all options), 8067 and
8082, and 6905.65 and 6883.30 (sum of element-wise multiplication of
individual conversions and revenue) respectively.

Fig.~\ref{fig:exp-2}~(bottom row) shows our posteriors of conversion rates,
revenue per visitor and gain per visitor for both the variants, using an
uniformed prior and Table~\ref{tab:agg-model} shows the experiment results.
Since we assumed that revenue follows exponential distribution, we
see that posteriors for revenue and gain have slightly shifted to the left in
the aggregated model when comparing Fig.~\ref{fig:exp-2}~(bottom row) and
Fig.~\ref{fig:exp-2}~(top row). The results from Table~\ref{tab:multi-model}
and Table~\ref{tab:agg-model} for conversion and revenue are
almost identical. The aggregated model shows more uncertainty in gain,
however the credible interval of improvement for the aggregate model is
inclusive of the credible interval of improvement for the multi-options model.

\section{Summary}
\label{sec:summary}

Decision making is central in running a business---with data-driven decisions
being the ones having the highest impact on output and
productivity~\cite{brynjolfsson+-2011}.
To support decision making, we (and many other companies) are continuously
running experiments. For easier interpretability we use a Bayesian approach
for the analysis of the experiments. In this paper, we introduced the three
most common scenarios in our company: The \textit{one option scenario} for
testing websites where the visitor has only the option to purchase one product.
The \textit{multi-option scenario} for testing websites where the visitor has
multiple products to choose from. And the \textit{aggregated scenario}, which
is a multi-option scenario but we only observe aggregated data. For all three
scenarios, we presented the Bayesian formulation of the models and how to
draw decisions based on sampling the estimated posteriors.

To showcase the presented approach in production, we showed a real-world
experiment for each of the scenarios. In the first experiment, we wanted to
find out the effect of different discounts. The analysis made us switch from
our baseline of 20\% discount to 40\% discount. We validated the results of
our models against the industry standard tool Google Optimize that also uses
black-box Bayesian inference. In the second experiment, we wanted to investigate
the nudging effect of progressive discounts. The result showed us that
nudging did not have an effect (probability to be the best based on revenue
is the same for both options). However, as the costs for showing the ads for the
progressive discounts where cheaper (probability to be the best based on gain),
we switched to the progressive discounts. The third experiment is based on the
second one, and showed that even with observing only aggregated data, we come
to the same conclusion.

\section*{Acknowledgment}

\nottoggle{anon}{
  We thank our collegues Alin Secareanu and Matthew Bonick for providing
  helpful comments during the writing process.
}{
  Anonymized acknowledgments.\\
}



\bibliography{idsc2020-prayas}{}
\bibliographystyle{IEEEtran}

\end{document}